\documentclass[11pt]{article} 
\usepackage{graphicx}
\usepackage{hyperref}
\usepackage{amssymb}
\usepackage{authblk}
\usepackage[caption=false]{subfig}

\date{\today}

\title{All-flavor search for a diffuse flux of cosmic neutrinos with 9 years of ANTARES data}

\begin{document}

\author[1]{A.~Albert}
\author[2]{M.~Andr\'e}
\author[3]{M.~Anghinolfi}
\author[4]{G.~Anton}
\author[5]{M.~Ardid}
\author[6]{J.-J.~Aubert}
\author[7]{J.~Aublin}
\author[7]{T.~Avgitas}
\author[7]{B.~Baret}
\author[8]{J.~Barrios-Mart\'{\i}}
% ...

\affil[1]{\scriptsize{GRPHE - Universit\'e de Haute Alsace - Institut universitaire de technologie de Colmar, 34 rue du Grillenbreit BP 50568 - 68008 Colmar, France}}
\affil[2]{\scriptsize{Technical University of Catalonia, Laboratory of Applied Bioacoustics, Rambla Exposici\'o, 08800 Vilanova i la Geltr\'u, Barcelona, Spain}}
\affil[3]{\scriptsize{INFN - Sezione di Genova, Via Dodecaneso 33, 16146 Genova, Italy}}
\affil[4]{\scriptsize{Friedrich-Alexander-Universit\"at Erlangen-N\"urnberg, Erlangen Centre for Astroparticle Physics, Erwin-Rommel-Str. 1, 91058 Erlangen, Germany}}
\affil[5]{\scriptsize{Institut d'Investigaci\'o per a la Gesti\'o Integrada de les Zones Costaneres (IGIC) - Universitat Polit\`ecnica de Val\`encia. C/  Paranimf 1, 46730 Gandia, Spain}}
\affil[6]{\scriptsize{Aix Marseille Univ, CNRS/IN2P3, CPPM, Marseille, France}}
\affil[7]{\scriptsize{APC, Univ Paris Diderot, CNRS/IN2P3, CEA/Irfu, Obs de Paris, Sorbonne Paris Cit\'e, France}}
\affil[8]{\scriptsize{IFIC - Instituto de F\'isica Corpuscular (CSIC - Universitat de Val\`encia) c/ Catedr\'atico Jos\'e Beltr\'an, 2 E-46980 Paterna, Valencia, Spain}}
% ...


\maketitle 



\begin{abstract}

 
The ANTARES detector is at present the most sensitive neutrino telescope in the Northern Hemisphere. The highly significant cosmic neutrino excess observed by the Antarctic IceCube detector can be studied with ANTARES, exploiting its complementing field of view, exposure, and lower energy threshold. Searches for an all-flavor diffuse neutrino signal, covering 9 years of ANTARES data taking, are presented in this letter. Upward-going events are used to reduce the atmospheric muon background. This work includes for the first time in ANTARES both track-like (mainly $\nu_\mu)$ and shower-like (mainly $\nu_e$) events in this kind of analysis. Track-like events allow for an increase of the effective volume of the detector thanks to the long path traveled by muons in rock and/or sea water. Shower-like events are well reconstructed only when the neutrino interaction vertex is close to, or inside, the instrumented volume. A mild excess of high-energy events over the expected background is observed in 9 years of ANTARES data in both samples. The best fit for a single power-law cosmic neutrino spectrum, in terms of per-flavor flux at 100 TeV, is $\Phi_0^{1f}(100\ \textrm{TeV}) = \left(1.7\pm 1.0\right) \times$10$^{-18}$\,GeV$^{-1}$\,cm$^{-2}$\,s$^{-1}$\,sr$^{-1}$ with spectral index $\Gamma = 2.4^{+0.5}_{-0.4}$. The null cosmic flux assumption is rejected with a significance of 1.6$\sigma$.

\end{abstract}


\maketitle



\textit{Introduction -} A diffuse flux of cosmic neutrinos -- here, and in the rest of this paper, the world \textit{neutrino} refers to both $\nu$ and $\bar{\nu}$ -- might originate from the ensemble of unresolved sources, too faint to be individually detected, and/or from the interactions of high-energy Cosmic Rays (CRs) as they propagate over cosmic distances. CRs can produce neutrinos when they inelastically interact on nucleons or photons. In the first case, the so-called $pp$ reaction, a large amount of secondaries is produced, including also short-lived mesons decaying into neutrinos \cite{bib:ahro}; the second case is described by photo-production processes, where the $p\gamma$ reaction produces a $\Delta$-resonance which gives pions. Neutrinos and $\gamma$-rays originate in the decay chain of these mesons \cite{bib:proto}. High energy photons, however, can interact with thermal protons and photons. These processes, in turn, create photons of lower energies, distorting the original $\gamma$-ray spectrum. Neutrinos are weakly interacting particles and consequently do not suffer from significant absorption processes due to the presence of matter and radiation fields. Thus, their spectral energy distribution is not degraded.

The observation of a diffuse flux of cosmic neutrinos, i.e.\ the measurement of its spectrum and flavor composition, would provide information on the CR production, acceleration and interaction properties. Neutrinos should follow the energy spectrum of their parent CRs, $\propto E^{-\Gamma}$ with $\Gamma\sim 2$ according to the diffusive shock acceleration mechanism \cite{bib:fermi}. Under the assumption that neutrinos are produced in charged meson decays, their flux at the source has a flavor composition as $\nu_e:\nu_\mu:\nu_\tau = 1:2:0$. Vacuum oscillations over cosmic distances produce equipartition in the three flavors at Earth. 

Searches for a diffuse flux of cosmic neutrinos by the IceCube collaboration have yielded the observation of an excess of events over the expected atmospheric background \cite{bib:ic, bib:ic_comb, bib:ICmu}. The measured flux can be modeled with a single power law $dN_\nu/dE_\nu=\Phi_0 E_\nu^{-\Gamma}$. Assuming an isotropic astrophysical neutrino flux at Earth in flavor equipartition, the best-fit spectral index is $\Gamma=2.50\pm 0.09$ and the normalization at 100 TeV is $\Phi^{3f}_0(100\ \textrm{TeV}) = 6.7^{+1.1}_{-1.2}\times 10^{-18}$\,GeV$^{-1}$\,cm$^{-2}$\,s$^{-1}$\,sr$^{-1}$ for an all-flavor flux ($3f$) \cite{bib:ic_comb}. The measurement of muon neutrinos coming only from the Northern Hemisphere yields a best-fit single-flavor flux $\Phi^{1f}_0(100\ \textrm{TeV}) = 9.0^{+3.0}_{-2.7}\times 10^{-19}$\,GeV$^{-1}$\,cm$^{-2}$\,s$^{-1}$\,sr$^{-1}$ and $\Gamma = 2.13\pm 0.13$ \cite{bib:ICmu}. The latter is sensitive to neutrinos of energies larger than 100 TeV because of the more abundant atmospheric background, while the former, all-flavor searches, have lower energy thresholds. The tension between the two measurements could hint at multiple cosmic contributions to the IceCube signal \cite{bib:anomaly}.

The ANTARES telescope \cite{bib:antares} can provide valuable information on the study of this signal, especially in the case of the presence of a Galactic contribution. Because of its lower energy threshold, ANTARES can constrain such a contribution, which is expected to be more intense at lower energies with respect to extragalactic signals. This applies both for point-like sources \cite{bib:ps} and for extended emission regions \cite{bib:gp,bib:gp_t}. Past searches for a diffuse flux of cosmic neutrinos with ANTARES data were below the sensitivity for detecting a signal at the level of the flux observed by IceCube \cite{bib:Anuflux, bib:Jutta_ICRC, bib:Flo}. The analysis presented in this letter is based on improved reconstruction techniques and, for the first time, on an all-flavor search. The livetime of the analyzed data sample is also largely extended with respect to previous analyses.

\textit{Detector and data sample -} The ANTARES underwater neutrino telescope \cite{bib:antares} is located 40\,km off-shore Toulon, France, in the Mediterranean Sea (42$^\circ$ 48$^\prime$ N, 6$^\circ$ 10$^\prime$ E). It consists of a three-dimensional array of 10-inch photomultiplier tubes (PMTs) distributed along 12, 450~m long, vertical lines, anchored to the sea-bed at a depth of about 2500\,m and kept taut by a top buoy.

 Neutrino detection is based on the observation of Cherenkov light induced in the medium by relativistic charged particles. Cherenkov photons can produce signals in the PMTs (``hits''). The position, time and collected charge of the hits are used to infer the direction and energy of the incident neutrino. Triggers based on combinations of local coincidences are applied to discard events produced by environmental light emitters like inorganic $^{40}$K decays and organic bioluminescence \cite{bib:antares_daq}.

Charged Current (CC) interactions of muon neutrinos produce a track signature in the detector. For these events, a median angular resolution as low as 0.4$^\circ$ is achieved \cite{bib:ps}. All-flavor Neutral Current (NC) as well as $\nu_e$ and $\nu_\tau$ CC interactions produce electromagnetic and hadronic showers with an almost point-like emission of Cherenkov photons. These events are reconstructed when the neutrino interacts close to, or inside, the instrumented volume \cite{bib:tino} with a median angular resolution of the order of 3$^\circ$ and a relative energy resolution as low as 10\% in the case of CC $\nu_e$ interactions above some tens of TeV.

Data collected from 2007 to 2015 are considered, corresponding to an equivalent livetime of 2450 days. In order to avoid biases in the optimization of the event selection, the analysis follows a {\it blind} policy, according to which all cuts are optimized on Monte Carlo only, with 10\% of the data used to check the agreement with simulations. Since the data acquisition conditions in the deep sea are variable, the simulation of the apparatus follows the data-taking conditions \cite{bib:rbr}.

\textit{Search method -} The search strategy follows a two-step procedure. First, an event selection chain is defined to overcome the large background of atmospheric muons. Second, since the cosmic signal has a harder energy spectrum and becomes more intense than the atmospheric one at high energies, an energy-related selection maximizes the sensitivity of the search. Thus, the Model Rejection Factor (MRF) procedure \cite{bib:MRF} based on the Feldman and Cousins upper limit estimation \cite{bib:F&C} is applied.

An isotropic flux over the whole sky is assumed for cosmic neutrinos, equally distributed into the three neutrino flavors and between $\nu$ and $\bar{\nu}$, and with single power law energy spectrum. Two possible spectral indexes are considered for the optimization of the event selection: $\Gamma=2.0$ and $\Gamma=2.5$.

The most abundant background comes from penetrating atmospheric muons reaching the apparatus. The MUPAGE simulation code \cite{bib:mupage1,bib:mupage2} is used to produce samples of atmospheric muon bundles deep underwater. The Earth can be used as a shield to reduce the influence of these muons, by discarding events that are reconstructed as downward-going. Nonetheless, a certain amount of atmospheric muon events could still be mis-reconstructed as upward-going. A selection based on the event reconstruction quality parameters (see below) allows for a significant reduction of their amount. 

Atmospheric neutrinos are produced together with muons and cannot be shielded by the Earth. Two spectral components contribute to the total atmospheric neutrino flux, namely the \textit{conventional} and the \textit{prompt} component. The former comes from neutrinos produced in the decays of pions and kaons; the latter originates from the decays of charmed hadrons. Charmed hadrons are much shorter-lived than pions and kaons and they almost immediately decay, producing a harder neutrino energy spectrum. In this work, the conventional component is described according to the calculations in Ref.\,\cite{bib:honda}, while the prompt component follows the prescription of Ref.\,\cite{bib:enberg}. Different assumptions in the description of atmospheric neutrino fluxes are accounted for as systematic effects.

Track-like events are reconstructed through a multi-step procedure based on likelihood fits \cite{bib:ps} using hits registered by the PMTs. The discrimination between downward-going atmospheric muons and high-energy neutrino induced events is accomplished by applying, \textit{a priori}, a cut on the reconstructed zenith angle ($\theta^{\textrm{track}}>90^\circ$). Neutrino induced events are then selected against wrongly reconstructed muons, keeping events with low angular error estimate and good track quality parameter \cite{bib:first_ps}. An energy-related variable such as the number of hits used in the track reconstruction is considered as well to reduce the number of background events. The resulting atmospheric neutrino rate in the track channel is about 1 event/day, with a contamination from atmospheric muons below the percent level in this sample.

\begin{figure}
  \centering
  \begin{subfloat}{\includegraphics[width=0.8\textwidth]{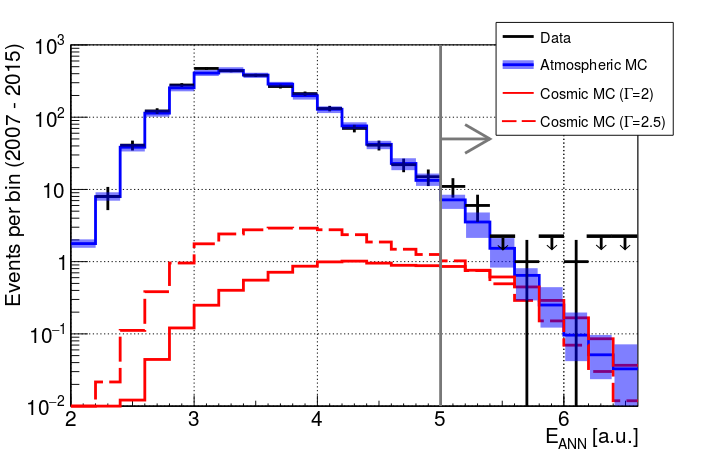}}
  \end{subfloat}\\
  \begin{subfloat}{\includegraphics[width=0.8\textwidth]{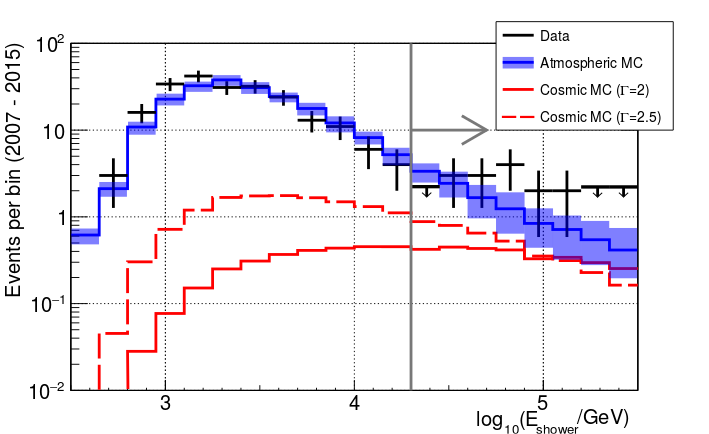}}
  \end{subfloat}
  \caption{Distribution of the energy estimator for track-like (top panel) and shower-like (bottom panel) events, after the event selection chain. The solid (dashed) red histogram shows the cosmic neutrino expectation for a cosmic flux proportional to E$^{-2}$ (E$^{-2.5}$) with normalization $\Phi_0^{1f}(100\ \textrm{TeV})=10^{-18}$ (1.5$\times$10$^{-18}$)\,GeV$^{-1}$\,cm$^{-2}$\,s$^{-1}$\,sr$^{-1}$. The blue line represents the sum of all atmospheric events, scaled up to match the fitted atmospheric contribution as described in the text. All the uncertainties related to this evaluation, taken into account as described in the text, are depicted as a shaded area. The gray line represents the energy-related cut. Data after unblinding are shown as black crosses. For empty bins, upper limits are indicated by a horizontal bar with an arrow beneath.}
  \label{fig:energy}
\end{figure}

Since the energy cannot be directly measured in the case of CC $\nu_\mu$ track-like events, a proxy for the event energy must be used. An algorithm based on Artificial Neural Networks \cite{bib:jutta} is used for this purpose. Figure~\ref{fig:energy} (top) shows the energy estimator distribution for the selected neutrino sample. The vertical line and arrow show the optimal selection cut obtained with the MRF procedure. Above the cut value $E_{ANN} > 5$, a total of 13.5 background neutrino events are expected, and about 3 -- 3.5 events should be produced by an IceCube-like signal. This depends on the spectral index and flux normalization resulting from the different best-fit values described above.
 
The event selection chain described in Ref.\,\cite{bib:tino} is used in this work to obtain a high-purity sample of shower-like events. No explicit veto on atmospheric events is applied; events reconstructed up to 10$^\circ$ above the horizon are considered and events entering into the final track-like sample are excluded. The muon-rejecting procedure leaves about 8$\times10^{-3}$ atmospheric muons per day in the sample. The resulting atmospheric neutrino rate in the shower channel is 0.1 events/day. However, the rejection of atmospheric muon events is more difficult at the highest energies, where the contamination is larger. The final selection is obtained after rejecting events with reconstructed shower energy below 20 TeV. This value arises from the MRF optimization to obtain the best sensitivity to the IceCube cosmic neutrino signal flux. The reconstructed shower energy distribution is shown in Fig. \ref{fig:energy} (bottom). After the energy-related selection, 10.5 background events (6.5 neutrinos and 4 atmospheric muons) are expected. Assuming a cosmic flux proportional to E$^{-2}$ (E$^{-2.5}$) with normalization $\Phi_0^{1f}(100\ \textrm{TeV})=10^{-18}$ (1.5$\times$10$^{-18}$)\,GeV$^{-1}$\,cm$^{-2}$\,s$^{-1}$\,sr$^{-1}$, compatible with the flux measured in Ref. \cite{bib:ic_comb,bib:ICmu}, 3 (3.5) signal events are expected.

\textit{Results -} The unblinding of the two samples yields a total of 33 events (19 tracks and 14 showers), as shown in Fig. \ref{fig:energy}. The expectation from Monte Carlo simulations for the background is 24 events (13.5 tracks and 10.5 showers), with an estimated uncertainty of $\pm7$ events. The uncertainties related to the atmospheric neutrino fluxes, to the detector efficiency and to the water properties have been accounted for. A $\pm25\%$ uncertainty on the normalization of the conventional atmospheric neutrino component is considered \cite{bib:atmo_unc}; the highest and lowest predictions from the computations of Ref.\,\cite{bib:enberg} have been used as uncertainty on the prompt component. It should be noted that IceCube measurements strongly constrain the predictions of prompt neutrino fluxes \cite{bib:ICmu}. The effect of the H3a \cite{bib:knee} model of the CR energy spectrum and composition around the \textit{knee} is also considered. 

Uncertainties on the background due to wrongly reconstructed atmospheric muons are taken into account, allowing for changes in the normalization by $\pm$40\% \cite{bib:atmomu}. The effect is negligible in the track channel, because of the low contamination in this sample, but represents around 50\% of the total uncertainty on the background for shower-like events.

Finally, the effect of the uncertainty on the detector response on signal and background is evaluated by varying input parameters in the Monte Carlo simulations. Water properties and the efficiency of the optical modules are varied according to the known uncertainties on the values used for the simulation. The corresponding effect is around 20\% at high energies and induces a change of shape in the energy estimation, accounted for in the fitting procedure presented below. The overall uncertainty on the background coming from all the effects mentioned above is of the order of $\pm$4 events for each sample. Uncertainties coming from independent sources are added in quadrature in the overall estimation, while correlated uncertainties are summed-up linearly.

Once these effects are taken into account, the observation can be translated into a 68\% confidence interval (C.I.) and a 90\% confidence level (C.L.) upper limit (U.L.) according to the method of Ref.\,\cite{bib:Conrad}. The results are reported in Table \ref{tab:limit}, together with the sensitivity of the analysis as estimated from the MRF procedure. These limits and sensitivities are valid in the energy range 40 TeV -- 7 PeV  (30 TeV -- 1.5 PeV) for spectral index $\Gamma=2.0$ $(2.5)$, where 90\% of the combined track and shower signal is expected. The observed excess does not translate into a significant observation of a cosmic signal, even though a null cosmic flux hypothesis can be excluded at 85\% C.L.\ .

\begin{table}
  \[
  \begin{array}{lcc}
    \hline
                                & \Gamma = 2.0        &  \Gamma = 2.5 \\
\hline
\Phi_0^{1f,\,90\% Sens.}(100\ \textrm{TeV})  &  1.2 \times10^{-18}  & 2.0 \times10^{-18} \\

\Phi_0^{1f,\,90\% U.L.}(100\ \textrm{TeV}) & 4.0\times10^{-18}    &  6.8\times10^{-18} \\

\Phi_0^{1f,\,68\% C.I.}(100\ \textrm{TeV}) & 0.29 - 2.9\times10^{-18} &  0.5 - 5.0\times10^{-18} \\

    \hline\hline
  \end{array}
  \]
  \caption{Sensitivity and unblinded results from counting statistics. The one-flavor 90\% confidence level sensitivity $\Phi_0^{1f,\,90\%Sens.}$ and upper limit $\Phi_0^{1f,\,90\% U.L.}$ flux normalization factors at 100 TeV are reported, as well as the 68\% confidence interval $\Phi_0^{1f,\,68\% C.I.}$, under the assumption of a cosmic spectrum proportional to E$^{-2}$ or E$^{-2.5}$. Systematic effects are included into these estimations. Fluxes are shown in units of GeV$^{-1}$\,cm$^{-2}$\,s$^{-1}$\,sr$^{-1}$.}
  \label{tab:limit}
\end{table}

\begin{figure}
  \centering
  \includegraphics[width=0.8\textwidth]{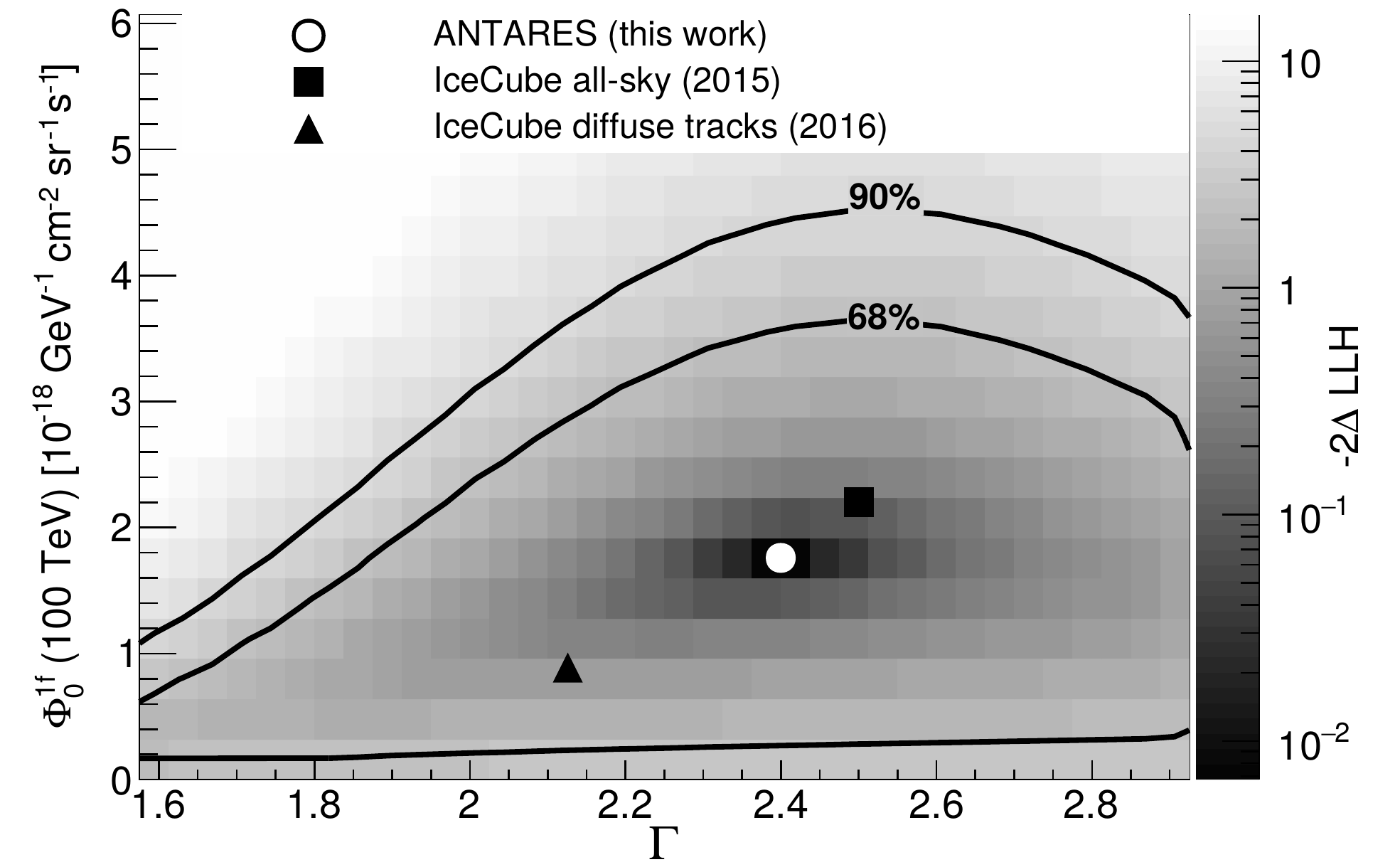}
  \caption{2D log-likelihood scan of the diffuse cosmic flux normalization and spectral index. The 68\% and 90\% confidence contours are shown as black lines. The empty circle is the best-fit point from this analysis, compared to the IceCube best fits from the all-sky combined analysis \cite{bib:ic_comb} (full square) and the diffuse flux analysis using tracks \cite{bib:ICmu} (full triangle). The color gradient represents the log-likelihood difference with respect to the best-fit point.}
  \label{fig:fit}
\end{figure}

The observed distributions of the energy estimators, after the final selection, are fitted using a maximum-likelihood method as done in Ref.\,\cite{bib:ic_comb}. Monte Carlo simulations are used to create templates of the cosmic signal and of the atmospheric backgrounds, considering different normalizations and spectral indexes for the signal. Binned distributions of energy estimators of data and of the Monte Carlo templates are considered. The final likelihood $L$ is given by the product of the individual likelihoods $L_{i,S}$ computed for each bin $i$ of the energy estimator distribution for the shower ($sh$) and track ($tr$) samples $S$ separately. The distribution of data and templates are compared considering Poisson statistics, with a Gaussian penalty factor to account for systematic effects on the Monte Carlo input parameters $\tau_j^*$:

\begin{eqnarray*}
\displaystyle L &= \prod_{S\in\{sh,tr\}}\prod_{i=0}^{N_{S}} L_{i,S} &\\
L_{i,S}&= e^{-\mu_{i,S}}\cdot\frac{\mu_i^{k_{i,S}}}{k_{i,S}!}\cdot \prod_j\frac{1}{2\pi\sigma[\tau_{i,S,j}]}e^{-\frac{\left(\tau_{i,S,j}-\tau_{i,S,j}^*\right)^2}{\sigma^2[\tau_{i,S,j}]}}&
\end{eqnarray*}
where $\mu_i$ is the expected number of events in the $i$-th bin from the simulated templates, $N_S$ is the number of bins in the energy estimator histogram for each event sample and $k_{i,S}$ is the number of events observed in data for that event sample in that bin. The nuisance parameters $\tau_{i,S,j}$ considered here are: the atmospheric neutrino background normalization; the residual atmospheric muon background, relevant in the shower analysis only; an energy-scale shift, which can be produced by the uncertainty on water properties and optical module efficiencies, as well as on the response of the PMTs. The considered effects modify the expected $\mu_i$ in the simulated templates analogously to what has been reported above. The atmospheric normalization is fitted simultaneously for the two samples, assuming that the background fluxes should follow the same modeling. The possible energy shift is considered separately for cascades and tracks. The 2D log-likelihood profile, after having fixed the nuisance parameters to the best-fit values, is shown in Fig. \ref{fig:fit}. The 68\% and 90\% C.L. contours from this analysis are shown together with the best-fit results from IceCube analyses. 

The likelihood profile shows a flat minimum region, which does not allow for constraining significantly the properties of the cosmic signal, but excludes extremely hard spectra or intense fluxes. The best-fit cosmic flux from ANTARES data yields a single-flavor normalization at 100 TeV of $\Phi_0^{1f}(100\ \textrm{TeV}) = \left(1.7\pm 1.0\right) \times 10^{-18}$\,GeV$^{-1}$\,cm$^{-2}$\,s$^{-1}$\,sr$^{-1}$ and a spectral index $\Gamma = 2.4^{+0.5}_{-0.4}$, when profiling the likelihood at the 68\% C.L.\ . The best-fit value for the atmospheric neutrino normalization is 25\% higher than the Monte Carlo simulations according to the predictions of Ref.\,\cite{bib:honda}, as also observed in Ref.\,\cite{bib:atmo}. This agrees with the event rates observed below the analysis energy threshold, where the sample is dominated by atmospheric events. An energy scale shift of $-$0.12 in the logarithm of the shower energy estimator provides the best-fit results; for the track energy estimator, a null shift is found. The same results are found when fitting the low-energy part of the energy distributions only, under the assumption that a cosmic signal would not be visible there.

Both IceCube best-fit points lie in the 68\% C.L. contour of this analysis. Not shown here are the 68\% C.L. contours from Ref.\,\cite{bib:ic_comb} and \cite{bib:ICmu}, which would also entirely be inside the 68\% C.L. contour depicted in Fig. \ref{fig:fit}. The hypothesis of a null cosmic flux is excluded at 1.6$\sigma$, assuming the best-fit hypothesis in a likelihood-ratio test. Even though this significance is not large, this result leans towards the observation of a cosmic neutrino flux compatible with the one observed in the IceCube data.


\textit{Acknowledgments -} The authors acknowledge the financial support of the funding agencies:
Centre National de la Recherche Scientifique (CNRS), Commissariat \`a
l'\'ener\-gie atomique et aux \'energies alternatives (CEA),
Commission Europ\'eenne (FEDER fund and Marie Curie Program),
Institut Universitaire de France (IUF), IdEx program and UnivEarthS
Labex program at Sorbonne Paris Cit\'e (ANR-10-LABX-0023 and
ANR-11-IDEX-0005-02), Labex OCEVU (ANR-11-LABX-0060) and the
A*MIDEX project (ANR-11-IDEX-0001-02),
R\'egion \^Ile-de-France (DIM-ACAV), R\'egion
Alsace (contrat CPER), R\'egion Provence-Alpes-C\^ote d'Azur,
D\'e\-par\-tement du Var and Ville de La
Seyne-sur-Mer, France;
Bundesministerium f\"ur Bildung und Forschung
(BMBF), Germany; 
Istituto Nazionale di Fisica Nucleare (INFN), Italy;
Nederlandse organisatie voor Wetenschappelijk Onderzoek (NWO), the Netherlands;
Council of the President of the Russian Federation for young
scientists and leading scientific schools supporting grants, Russia;
National Authority for Scientific Research (ANCS), Romania;
Mi\-nis\-te\-rio de Econom\'{\i}a y Competitividad (MINECO):
Plan Estatal de Investigaci\'{o}n (refs. FPA2015-65150-C3-1-P, -2-P and -3-P, (MINECO/FEDER)), Severo Ochoa Centre of Excellence and MultiDark Consolider (MINECO), and Prometeo and Grisol\'{i}a programs (Generalitat
Valenciana), Spain; 
Ministry of Higher Education, Scientific Research and Professional Training, Morocco.
We also acknowledge the technical support of Ifremer, AIM and Foselev Marine
for the sea operation and the CC-IN2P3 for the computing facilities.

\end{document}